\documentclass[12pt]{article}  
  
\usepackage{epsfig,epsf}  
\usepackage{graphics}  
\usepackage{cite}  
\usepackage{a4wide}  
\usepackage{xcolor} 

\usepackage{pstricks} 
 
\usepackage{amsmath}  
\usepackage{amsthm}  
\usepackage{amsfonts}  
\usepackage{amssymb}  
\usepackage{slashed}  
  
\renewcommand{\thefootnote}{\fnsymbol{footnote}}  
  
\newcommand{\eps}{\epsilon}

\newcommand{\be}{\begin{equation}}  
\newcommand{\ee}{\end{equation}}  
\newcommand{\ba}{\begin{eqnarray}}  
\newcommand{\ea}{\end{eqnarray}}  
\newcommand{\baa}{\begin{eqnarray*}}  
\newcommand{\btab}{\begin{tabular}}  
\newcommand{\etab}{\end{tabular}}  
\newcommand{\eaa}{\end{eqnarray*}}  
  
\newcommand \ket [1] {|{#1}\rangle}  
\newcommand \bra [1] {\langle {#1}|}

\newcommand{\qbar} {\overline{q}}  
\newcommand{\rd}{\,\mathrm{d}}

\newcommand{\nn}{\nonumber}    
\newcommand{\ubar}{\overline{u}}

\def \eps {\epsilon}  
\def \gln {\gamma\ell\nu}  
\def \gf {\gamma_5}  
  
\def \rd {\!{\rm d}}  
  
\numberwithin{equation}{section}  
  
\begin{document}  
  
\allowdisplaybreaks  
\thispagestyle{empty}  
  
\begin{flushright}  
{\small  
TTK-11-51\\  
SFB/CPP-11-56\\  
October 14, 2011   
}  
\end{flushright}  
  
\vskip1.5cm  
\begin{center}  
\textbf{\Large\boldmath $B$ meson distribution  
amplitude from $B\to \gln$ 
}  
\\  
\vspace{1.2cm}  
{\sc M.~Beneke} 
and {\sc J.~Rohrwild}  
\\[0.5cm]  
\vspace*{0.1cm} {\it  
Institut f\"ur Theoretische Teilchenphysik und Kosmologie,\\   
RWTH Aachen University, \\   
D-52074 Aachen, Germany}   
  
\def\thefootnote{\arabic{footnote}}  
\setcounter{footnote}{0}  
  
\vskip2cm  
\textbf{Abstract}\\  
\vspace{1\baselineskip}  
\parbox{0.9\textwidth}{   
We reconsider the utility of the radiative decay $B\to \gln$ with  
an energetic photon in the final state for determining parameters  
of the $B$-meson light-cone distribution amplitude.   
Including  $1/m_b$ power corrections and radiative corrections at  
next-to-leading logarithmic order, we perform an improved   
analysis of the existing BABAR data. We find a provisional lower  
limit on the inverse moment of the  $B$ meson distribution  
amplitude, $\lambda_B$, which, due to the inclusion of  
radiative and power corrections, is significantly lower than the  
previous result. More data with large photon energy is, however,  
required to obtain reliable results, as should become available  
in the future from SuperB factories.}  
  
\end{center}  
  
  
\newpage  
\setcounter{page}{1} 
  
 
\section{Introduction}  
  
The decay of the charged $B$ meson into a photon, lepton  
and neutrino is sometimes perceived as an unwanted background to  
the purely leptonic decay  
process $B^-\to \ell \bar \nu$~\cite{Becirevic:2009aq}, which allows  
for a determination of $V_{ub}$. Still, the radiative leptonic decay   
is of interest in itself for the theory of heavy meson decays,  
especially when the energy $E_\gamma$ of the photon is of order of  
the bottom quark mass, $m_b$. Its factorization properties have  
been studied at leading order in the heavy-quark expansion  
\cite{Korchemsky:1999qb,DescotesGenon:2002mw,Lunghi:2002ju,Bosch:2003fc} 
and it has been shown that the decay amplitude can be calculated  
in terms of the inverse and inverse-logarithmic moments of  
the $B$-meson light-cone distribution  
amplitude~\cite{Beneke:1999br,Grozin:1996pq,Beneke:2000wa}.  
 
The branching fraction of the radiative decay depends very  
strongly on the inverse moment $\lambda_B$, where $1/\lambda_B =  
\int_0^\infty \rd\omega\,\Phi_{B+}(\omega)/\omega$. It therefore  
seems very well suited as an observable to measure $\lambda_B$,  
which is an important parameter in the QCD factorization  
approach to non-leptonic $B$ decays~\cite{Beneke:1999br},  
but is very difficult to obtain reliably by theoretical  
methods, the most advanced being QCD sum  
rules~\cite{Braun:2003wx}. We are aware of only two analyses  
by the BABAR collaboration that set limits on the $B^-\to\gamma \ell\bar\nu$  
branching fraction and 
$\lambda_B$~\cite{Aubert:2007yh,Aubert:2009ya}. The first  
reports $\lambda_B > 669\,\mbox{MeV} \,(591\,\mbox{MeV})$  
(depending on the treatment of priors), while the second,   
published analysis concludes the significantly weaker  
limit $\lambda_B > 300\,\mbox{MeV}$. The first result would be  
rather troublesome for non-leptonic $B$ decay phenomenology,  
which needs $\lambda_B\approx 200\,\mbox{MeV}$ to achieve  
a satisfactory description of color-suppressed decay  
modes~\cite{Beneke:2005vv,Bell:2009fm,Beneke:2009ek}. 
 
The BABAR analyses should be taken with a grain of salt, since,  
presumably in order not to sacrifice statistics, they do not require  
the photons to be sufficiently energetic for the theoretical  
calculation to be valid. This can certainly be improved in  
the future, in particular with the high statistics foreseen at  
the SuperB experiments. They also do not include radiative  
corrections, which is one of our concerns in this note. We show  
that after including next-to-leading logarithmically  
resummed corrections, and after correcting an error in the  
literature in the leading $1/m_b$ correction, the predicted  
branching fraction is significantly smaller. This reduces the  
lower limit on $\lambda_B$ considerably. 
 
The outline this paper is as follows. In Sec.~\ref{definition:sec}  
and the Appendix we briefly review the theoretical background and  
summarize the expression for the $B\to\gamma\ell\nu$ amplitude.  
Sec.~\ref{FormFactors:sec}  
discusses the size and stability of radiative corrections  
and the $B\to\gamma$ form factors themselves.  
In Sec.~\ref{Experiment:sec} we repeat the BABAR analysis  
in order to demonstrate the impact of our results on the  
bound on $\lambda_B$. We  conclude in Section~\ref{conl:sec}.

  
\section{Theory summary of $B\to\gln$ decay}  
\label{definition:sec}  
 
We consider the decay of a $B$ meson with mass $m_B$ and  
momentum $p^\mu =m_B v^\mu$ into a  
photon with momentum $q$, a neutrino with momentum $p_\nu$ and  
a lepton (momentum $p_\ell$). The lepton and neutrino are assumed   
to be massless, which restricts us to $\ell=e,\mu$. In the $B$  
meson rest frame the photon energy satisfies $E_\gamma\leq m_B/2$.  
We introduce the abbreviations $x_i=\frac{2 E_i}{m_B}$ where   
$i$  can be either $\gamma$, $\ell$ or $\nu$. We  
have $0\leq x_i\leq 1$ and $x_\gamma+x_\ell+x_\nu=2$.  
 The amplitude for the decay $B\to \gln$ can be written as    
\begin{equation} 
\label{amplitude}  
\mathcal{A}(B^-\to\gamma\ell\bar\nu) = \frac{G_F V_{ub}}{\sqrt{2}}  
\,\bra{\ell \bar \nu \gamma} \bar\ell \gamma^{\mu}(1-\gf)\nu  \cdot  
\ubar \gamma^{\mu}(1-\gf) b \ket{B^-} \;.  
\end{equation}  
The photon can be emitted either from the final-state lepton or  
from one of the constituents of the $B$ meson.  This can be made explicit   
by rewriting the matrix element using the electromagnetic current  
$j^\mu_{\rm em}=\sum_{q}Q_q\qbar\gamma_{\mu}q+Q_\ell\bar\ell\gamma_{\mu} 
\ell$. To first order in electromagnetic and to all orders in  
the strong interaction, we have       
\begin{eqnarray}  
&& \bra{\ell \bar \nu \gamma} \overline{\ell} \gamma^{\mu}(1-\gf)\nu\cdot   
 \ubar \gamma_{\mu}(1-\gf) b \ket{B^-}=  
\nn \\  
&& \hspace*{1cm} 
= -ie\eps^\star_\nu  
\left[\bra{\ell \bar \nu}\bar\ell\gamma^{\mu}(1-\gf)\nu) \ket{0} 
\cdot\int\rd ^4 x \,e^{iqx}   
\bra{0} {\rm T}\lbrace j_{\rm em}^\nu(x)  
(\ubar \gamma_{\mu}(1-\gf) b)(0) \rbrace \ket{B^-} \right. 
\qquad \nn \\  
&& \hspace*{1cm}  
\left.\qquad\qquad+\int \rd^4 x\, e^{iqx} \bra{\ell\bar \nu }  
{\rm T} \lbrace j_{\rm em}^\nu(x)  
(\bar \ell \gamma^{\mu}(1-\gf)\nu)(0) \rbrace \ket{0}  
\cdot \bra{0}\ubar \gamma_{\mu}(1-\gf)b\ket{B^-}\right]  
\nn \\[0.2cm] 
&&\hspace*{1cm}  
= e\eps^\star_\nu \,\bar u_{\ell} \gamma_{\mu}(1-\gf)u_\nu  
\cdot T^{\nu\mu}(p,q) -  i e Q_\ell f_B \cdot \bar u_\ell  
\slashed{\eps}^\star (1-\gf) u_\nu \,. 
\label{amplitudeviacurrent}  
\end{eqnarray}  
Note that we use $iD^\mu=i\partial^\mu - Q_\psi e A^\mu_{em}$ for the  
QED covariant derivative with $e$ the charge of the positron, and  
$Q_\psi$ the electric charge of fermion $\psi$ in units of $e$.  
The first term in the above equation corresponds to the emission from  
the meson constituents whereas the second term describes the emission  
from the lepton, and can be calculated exactly using 
\begin{equation} 
\bra{0}\ubar \gamma^{\mu}(1-\gf)b\ket{B^-(p)}=-i f_B p^\mu\,. 
\end{equation}     
The hadronic tensor can be parameterized as  
\begin{eqnarray} 
T_{\nu\mu}(p,q) &=& (-i) \int\rd ^4 x \,e^{iqx}   
\bra{0} {\rm T}\lbrace j_{\nu, \rm em}(x)  
(\ubar \gamma_{\mu}(1-\gf) b)(0) \rbrace \ket{B^-}  
\nn \\ 
&&\hspace*{-1.8cm}  
= \,  (-i) \left[i \epsilon_{\mu\nu\rho\sigma} v^\rho q^\sigma\, 
F_V(E_\gamma) + (g_{\mu\nu} v\cdot q-v_\nu q_\mu)\,\hat F_A(E_\gamma)  
+\frac{v_\nu v_\mu}{v\cdot q}\,f_B m_B + q_\nu\mbox{-terms} 
\right].\qquad 
\label{hadtensor} 
\end{eqnarray} 
The terms proportional to $q_\nu$ are irrelevant, since  
$\epsilon^\star\cdot q=0$. The $v_\nu v_\mu$ structure is often referred  
to as ``contact term''. Its coefficient is fixed by the 
electromagnetic current conservation Ward identity  
$q_{\nu}T^{\nu\mu}=-i f_B p^\mu$  
\cite{Khodjamirian:2001ga}\footnote{The sign difference compared to  
this reference is due to our different convention for the 
electromagnetic covariant derivative.}.  
The remainder consists of  
two form factors. In the following we shall describe the QCD  
calculation of these form factors for photon energies  
of order (but not necessarily near) $m_B/2$. 
 
With the help of $(p-q)^\mu \,\bar u_{\ell} \gamma_{\mu}(1-\gf)u_\nu 
=0$, valid for massless leptons, we may replace 
\begin{equation} 
 (g_{\mu\nu} v\cdot q-v_\nu q_\mu)\,\hat F_A(E_\gamma)  
+\frac{v_\nu v_\mu}{v\cdot q}\,f_B m_B \to  
 (g_{\mu\nu} v\cdot q-v_\nu q_\mu)\,F_A(E_\gamma)  
+ g_{\mu\nu} f_B 
\label{id1}  
\end{equation} 
in (\ref{hadtensor}), where the new axial form factor is defined as  
\begin{equation} 
F_A = \hat F_A + \frac{Q_\ell f_B}{E_\gamma}\,. 
\label{faredef} 
\end{equation} 
In this form the $g_{\mu\nu} f_B$ term in (\ref{id1}) cancels  
precisely the last term in (\ref{amplitudeviacurrent}) from  
photon emission off the lepton, and the amplitude (\ref{amplitude})  
is expressed entirely in terms of the two form factors  
$F_{V,A}$. We use this convention below.  
However, the decomposition (\ref{hadtensor}) is useful for 
calculations, since it allows us to assume that the indices  
$\mu,\nu$ are transverse relative to the four-vectors $v$ and  
$q$, such that $F_V$ and $\hat F_A$ can be extracted from the  
$\epsilon_{\mu\nu\rho\sigma}$ and $g_{\mu\nu}$ structures of the  
hadronic tensor, respectively.  
 
Squaring the amplitude, the doubly differential decay width in the  
$B$ rest frame reads  
\begin{equation} 
\frac{\rd^2\Gamma}{\rd E_\gamma \,\rd E_\ell}= 
\frac{\alpha_{\rm em}G_F^2 |V_{ub}|^2}{16 \pi^2} 
m_B^3(1-x_\gamma)\left[(1-x_\nu)^2(F_A+F_V)^2 + 
  (1-x_\ell)^2(F_A-F_V)^2  \right], 
\label{doublediff} 
\end{equation} 
where $E_\ell+E_\gamma\geq m_B/2$, and the form factors depend on  
$E_\gamma$ but not on the lepton energy $E_\ell$. Further  
integration results in  
\begin{equation} 
\frac{\rd\Gamma}{\rd E_\gamma}= 
\frac{\alpha_{\rm em}G_F^2 |V_{ub}|^2}{48 \pi^2} 
m_B^4 (1-x_\gamma)x_\gamma^3  
\left[F_A^2+F_V^2\right]. 
\end{equation} 
For energetic photons, the form factors are given by  
\begin{eqnarray} 
F_V(E_\gamma) &=& \frac{Q_u m_B f_B}{2 E_\gamma\lambda_B(\mu)} \,  
R(E_\gamma,\mu) +  
\left[\xi(E_\gamma) +  
\frac{Q_b m_B f_B}{2 E_\gamma m_b} +  
\frac{Q_u m_B f_B}{(2 E_\gamma)^2}   
\right], 
\nn \\[0.2cm] 
F_A(E_\gamma) &=& \frac{Q_u m_B f_B}{2 E_\gamma\lambda_B(\mu)} \,  
R(E_\gamma,\mu) +  
\left[\xi(E_\gamma) -  
\frac{Q_b m_B f_B}{2 E_\gamma m_b} -  
\frac{Q_u m_B f_B}{(2 E_\gamma)^2} +   
\frac{Q_\ell f_B}{E_\gamma}   
\right].\quad 
\label{ffs} 
\end{eqnarray} 
The first term represents the leading-power contribution in the  
heavy-quark expansion with $R(E_\gamma,\mu)$ a radiative correction  
factor that equals one at tree level. Note that this term  
is the same for the vector and axial form  
factor~\cite{Korchemsky:1999qb,DescotesGenon:2002mw,Lunghi:2002ju,Bosch:2003fc}. 
The terms in square brackets  
are $1/m_b$ power corrections relative to the leading term.  
They consist of a term $\xi(E_\gamma)$ that is common to both form  
factors (``symmetry-preserving'') and other terms of a simple form 
that differ (``symmetry-breaking'').  We do not include  
perturbative radiative corrections to the power-suppressed terms. 
 
We note that $(F_V-F_A)/(F_V+F_A)\sim \Lambda_{\rm QCD}/m_b$ is  
suppressed in the heavy-quark limit due to helicity conservation.  
The second term proportional to $(1-x_\ell)^2$ in (\ref{doublediff})  
is therefore 
suppressed. 
 
\subsubsection*{Radiative corrections}  
 
Radiative corrections to the $B^-\to\gamma\ell\bar\nu$ process were 
first calculated in a $k_t$-dependent 
approach~\cite{Korchemsky:1999qb}; see \cite{Charng:2005fj} for an 
extended analysis of the decay in this approach. However, there is no need not to  
integrate over $k_t$. The all-order 
factorization formula~\cite{Lunghi:2002ju,Bosch:2003fc} refers  
to this situation. The one-loop radiative corrections in collinear  
factorization have been computed some time  
ago~\cite{DescotesGenon:2002mw,Lunghi:2002ju,Bosch:2003fc} and we 
summarize them here. Our improvement consists in completing the  
next-to-leading logarithmic (NLL) summation of logarithms of  
$m_b/\Lambda_{\rm QCD}$, since the two-loop anomalous dimension  
of the heavy-light current is now known. 
 
The radiative correction can be written as a product of several factors, 
\begin{equation} 
\label{rfact} 
R(E_\gamma,\mu) = C(E_\gamma,\mu_{h1})K^{-1}(\mu_{h2})\times  
U(E_\gamma,\mu_{h1},\mu_{h2},\mu) \times J(E_\gamma,\mu), 
\end{equation} 
which come from the different scales contributing to the process. The 
multiplicative structure becomes transparent if the decay is analyzed 
in soft-collinear effective theory  
(SCET)~\cite{Bauer:2000yr,Bauer:2001yt,Beneke:2002ph,Beneke:2002ni}  
as done in~\cite{Lunghi:2002ju,Bosch:2003fc}. The first factor  
arises from the hard scale $m_b$ when the QCD heavy-to-light  
current is matched to the corresponding SCET current: 
\begin{equation} 
\ubar\gamma_{\mu_\perp} (1-\gf) b = C(E_\gamma,\mu) \, 
\bar\xi W_c \gamma_{\mu_\perp} (1-\gf) h_v + \ldots 
\label{scet1match} 
\end{equation} 
with~\cite{Bauer:2000yr} 
\begin{equation} 
\label{hardPart} 
C(E_\gamma,\mu) = 1 +  
\frac{\alpha_s C_F}{4\pi}\left(-2 \ln^2 \frac{2 E_\gamma}{\mu}  
+ 5 \ln \frac{2 E_\gamma}{\mu}  -\frac{3-2 x}{1-x} \ln 
  x - 2 {\rm Li}_2(1-x) -6 -\frac{\pi^2}{12} \right) 
\end{equation} 
and $x=2 E_\gamma/m_b$. 
The two-loop correction is also  
known~\cite{Bonciani:2008wf,Asatrian:2008uk,Beneke:2008ei,Bell:2008ws}, 
but at NLL accuracy we only need the one-loop term and two-loop 
anomalous dimension of the SCET current, which can be inferred  
from~\cite{Bonciani:2008wf,Asatrian:2008uk,Beneke:2008ei,Bell:2008ws}. 
After inserting (\ref{scet1match}) into (\ref{hadtensor}),  
the hadronic tensor factorizes into a hard-collinear contribution  
from the scale $(m_b\Lambda_{\rm QCD})^{1/2}$ and  
moments of the non-perturbative light-cone distribution amplitude (LCDA)   
of the $B$ meson. We define these moments as 
\begin{equation} 
\frac{1}{\lambda_B(\mu)} = \int_0^\infty\frac{\rd\omega}{\omega}\, 
\Phi_{B+}(\omega,\mu), 
\qquad 
\sigma_n(\mu) = \lambda_B(\mu)\int_0^\infty\frac{\rd\omega}{\omega}\, 
\ln^n\frac{\mu_0}{\omega}\,\Phi_{B+}(\omega,\mu) 
 \end{equation} 
where $\mu_0=1\,\mbox{GeV}$ is a fixed reference scale which is  
part of the definition of the inverse-logarithmic 
moments.\footnote{Note the difference with~\cite{Braun:2003wx},  
which sets $\mu_0\to\mu$. Our definition avoids the appearance of a 
large logarithm when $\sigma_n$ is evolved to the hard-collinear  
scale and features  $\rd\sigma_n/\,\rd\ln\mu = {\cal O}(\alpha_s)$.} 
The hard-collinear radiative correction  
reads~\cite{Lunghi:2002ju,Bosch:2003fc} 
\begin{equation} 
\label{hcPart} 
J(E_\gamma,\mu) = 1 +  
\frac{\alpha_s C_F}{4\pi}\left( 
\ln^2\frac{2 E_\gamma \mu_0}{\mu^2}  
-2 \sigma_1(\mu)  \ln\frac{2 E_\gamma \mu_0}{\mu^2} 
-1 -\frac{\pi^2}{6} + \sigma_2(\mu) 
\right). 
\end{equation} 
Since the SCET current in (\ref{scet1match}) uses the static 
heavy-quark field $h_v$, we encounter the static $B$ meson decay constant  
when taking the matrix element. We re-express this in terms of the  
QCD decay constant $f_B$, which introduces the conversion factor  
\begin{equation} 
\label{fbfactor} 
K(\mu) = 1 + \frac{\alpha_s C_F}{4\pi}  
\left( \frac32\ln\frac{m_b^2}{\mu^2}-2 \right). 
\end{equation} 
 
Inspecting (\ref{hardPart}), (\ref{hcPart}) and (\ref{fbfactor})  
shows that there is no common value of $\mu$ that avoids  
parametrically large logarithms of order $\ln m_b/\mu_0$.  
These logarithms can be summed to all orders  
by solving a renormalization group  
equation~\cite{Bosch:2003fc}, which introduces the evolution factor  
$U(E_\gamma,\mu_{h1},\mu_{h2},\mu)$ into (\ref{rfact}).  
Its explicit expression is given in the appendix. The  
hard scales $\mu_{h1},\mu_{h2}$ can (and should) now be taken  
${\cal O}(m_b)$, the hard-collinear scale  
$\mu\sim {\cal O}(m_b\Lambda_{\rm QCD})^{1/2}$.  
Eqs.~(\ref{hardPart}) and (\ref{fbfactor}) suggest that  
the hard scale $\mu_{h1}$ is $2 E_\gamma$ rather than $m_b$,  
while $\mu_{h2}\sim m_b$, which motivates keeping the two  
hard scales distinct in the general expressions. However,  
we might also set them equal,  
which is the conventional procedure. 
 
\subsubsection*{Power corrections}  
 
\begin{figure} 
\begin{center} 
\includegraphics[width=9cm]{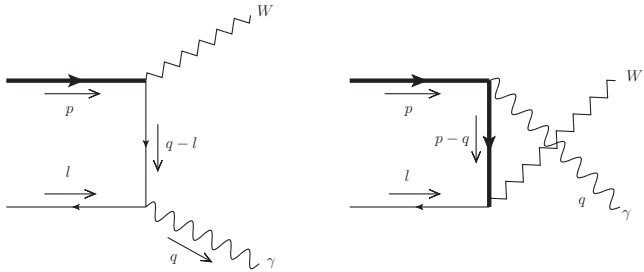}  
\end{center} 
\caption{Leading-order diagrams. The left graph shows the 
leading-power contribution from photon emission from the up   
anti-quark. Emission from the heavy $b$-quark (right) is  
power suppressed.}  
\label{LOdiagrams}  
\end{figure}  
 
For phenomenology $1/m_b$ corrections are presumably important.  
Power corrections 
are notoriously difficult for factorization approaches, but  
$B\to\gamma \ell\nu$ is arguably the simplest environment to study  
them.\footnote{See \cite{Beneke:2003pa} for a discussion of  
$B\to\gamma \ell\nu$ at order $1/m_b$.} Here we explain the origin  
of the ``tree-level'' $1/m_b$-suppressed terms in (\ref{ffs}),  
which are the most relevant to phenomenology, and defer the general  
discussion of $1/m_b$ power corrections to future work \cite{bhr}. 
 
The two diagrams for the tree-level $b\bar u\to \gamma W^*$ amplitude  
are shown in Fig.~\ref{LOdiagrams}. Since the spectator-quark  
momentum $l\sim\Lambda_{\rm QCD}$ is soft, the propagator joining  
the $W$ and $\gamma$ lines has hard-collinear  
virtuality $(q-l)^2 = 2q\cdot l  
\sim m_b \Lambda_{\rm QCD}$, when the photon is emitted from the  
up anti-quark (left), but hard virtuality $(p-q)^2 \sim m_b^2$  
in case of emission from the heavy quark (right). For this  
reason only emission from the light anti-quark is responsible for  
the leading-power contribution in (\ref{ffs}). The emission  
from the heavy quark can easily by calculated and results in  
the power-suppressed term proportional to the bottom-quark charge $Q_b$ in   
(\ref{ffs}). The term proportional to $Q_\ell$ present only in $F_A$  
is the contribution from emission off the lepton,  
see (\ref{faredef}).  
 
The remaining two terms in square brackets  
in (\ref{ffs}) come from power corrections to the emission  
off the light anti-quark. To understand the form of these terms,  
we consider the intermediate light-quark propagator  
(see figure) 
\begin{equation}  
\frac{i(\slashed{q}-\slashed{l})}{(q-l)^2} =  
-\frac{i \slashed{q}}{2q\cdot l} +  
\hspace*{-0.65cm}  \underbrace{\frac{i \slashed{l}}{2q\cdot l}}_{ 
\text{power suppressed}} \hspace*{-0.65cm} 
= - \frac{i \slashed{n}_-}{4 l_-} 
+  \left[\frac{i l_+\slashed{n}_-}{4E_\gamma l_-} 
+ \frac{i \slashed{l}_\perp}{2E_\gamma l_-} 
+ \frac{i \slashed{n}_+}{4E_\gamma}\right], 
\label{hcprop} 
\end{equation}   
using $q^2=l^2=0$. We also express $q^\mu=E_\gamma n_-^\mu$,  
$l^\mu = l_+ n_-^\mu/2+  
l_- n_+^\mu/2 +l_\perp^\mu$ in terms of two light-like vectors $n_\pm^\mu$ with  
$n_+ \cdot n_-=2$, spanning the plane of $q$ and $v$. The first two  
terms in square brackets are non-local. Before integrating out the  
hard-collinear scale they are exactly reproduced by time-ordered  
products of currents with SCET interactions. They may be matched to  
sub-leading $B$-meson distribution amplitudes, but it is not evident 
that this can be done without encountering endpoint  
divergences \cite{Beneke:2003pa}.  
The important point here is that it can be shown \cite{bhr}  
that these terms are symmetry-preserving, i.e.~they contribute  
equally to the vector and axial form factors. Hence we introduce  
a function $\xi(E_\gamma)$ in (\ref{ffs}) to parameterize  
this unknown contribution. The last term in (\ref{hcprop}) is a  
local term that contributes with opposite sign to the two form 
factors. Being local, it can be expressed through $f_B$ and  
yields the remaining term proportional to $Q_u$ in (\ref{ffs}).  
Numerically, this contribution is larger than emission from the heavy  
quark, due to its  
enhancement for smaller photon energies and  
the larger electric charge of the up quark. 
  
Tree-level power corrections have been computed previously   
\cite{Korchemsky:1999qb}, but the emission from the lepton and  
the sub-leading term from emission from the light anti-quark 
have been missed in this work. Also the contribution from  
emission from the heavy quark to $F_V$ has an incorrect sign,  
and appears as symmetry-preserving rather than -breaking. The  
difference is important numerically.

\section{Impact of radiative and power corrections} 
\label{FormFactors:sec} 
 
\begin{table} 
\begin{center} 
\begin{tabular}{c|c||c|c} 
 
parameter & value & parameter & value \\ 
\hline\hline  
&&&\\[-0.3cm] 
$G_F$ &  $1.16637\cdot 10^{ - 5} \,{\rm GeV}^{ - 2}$ & 
$\tau_{B_d}$  & $1.64\cdot 10^{-12}\, \rm s$ \\ 
$\alpha_{em}$ & $1/129$ &  
$f_B\,[\mbox{MeV}]$ &  $195\pm 10$ \\  
$\Lambda_{\overline{\rm MS}}^{n_f=4}$ & $289.9\,\mbox{MeV}$ & 
$m_b\,[\mbox{GeV}]$ &  $ 4.8\pm 0.1$ \\ 
$m_B$ &  $5279\,\rm MeV$ & 
$\lambda_B(1\,\mbox{GeV})$ & $ 350\,\mbox{MeV}$ \\ 
$|V_{ub}|_{\,\text{incl.}} $  &  $4.27 \cdot 10^{-3}$    & 
$\sigma_1(1\,\mbox{GeV})$ & $1.5\pm 1$ \\ 
$|V_{ub}|_{\,\text{excl.}}$  &  $3.38 \cdot 10^{-3}$ &  
$\sigma_2(1\,\mbox{GeV})$ & $3\pm 2$ \\ 
 \end{tabular} 
\caption{Central values and ranges of the input parameters.   
The four-flavour $\Lambda$ parameter  
  corresponds to $\alpha_s(M_Z) = 
  0.1185$ with decoupling of the bottom quark at the scale $m_b$. 
\label{InputData} } 
\end{center} 
\end{table} 
 
In this section we discuss the size of radiative and power corrections  
and the theoretical uncertainty attached to the form-factor  
calculation. The Standard Model and $B$-meson parameters that we use  
here and below in the computation of the differential branching fraction  
are listed in Tab.~\ref{InputData}. 
 
\begin{figure}[t] 
\begin{center} 
\includegraphics[width=.48\textwidth]{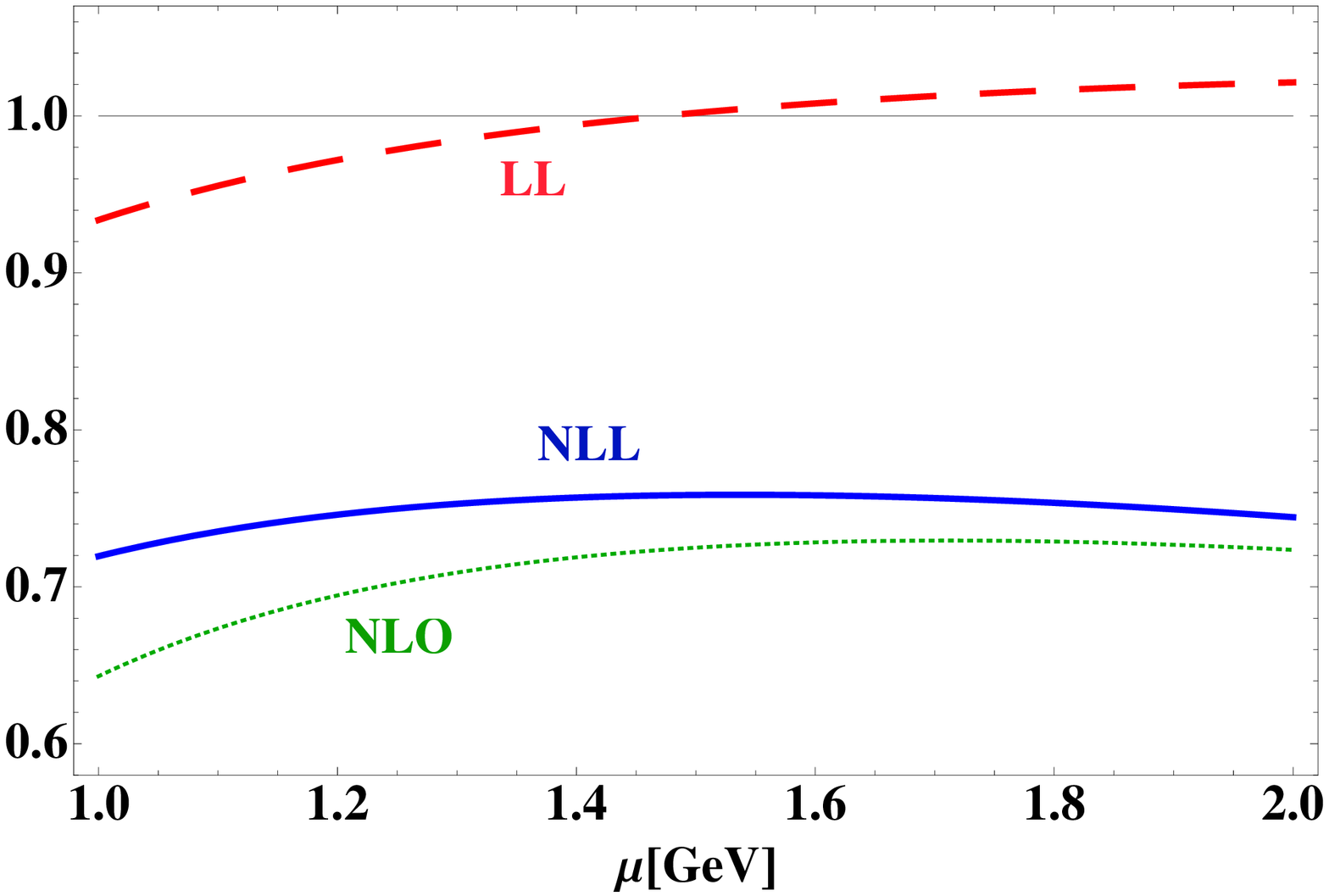}   
\includegraphics[width=.50\textwidth]{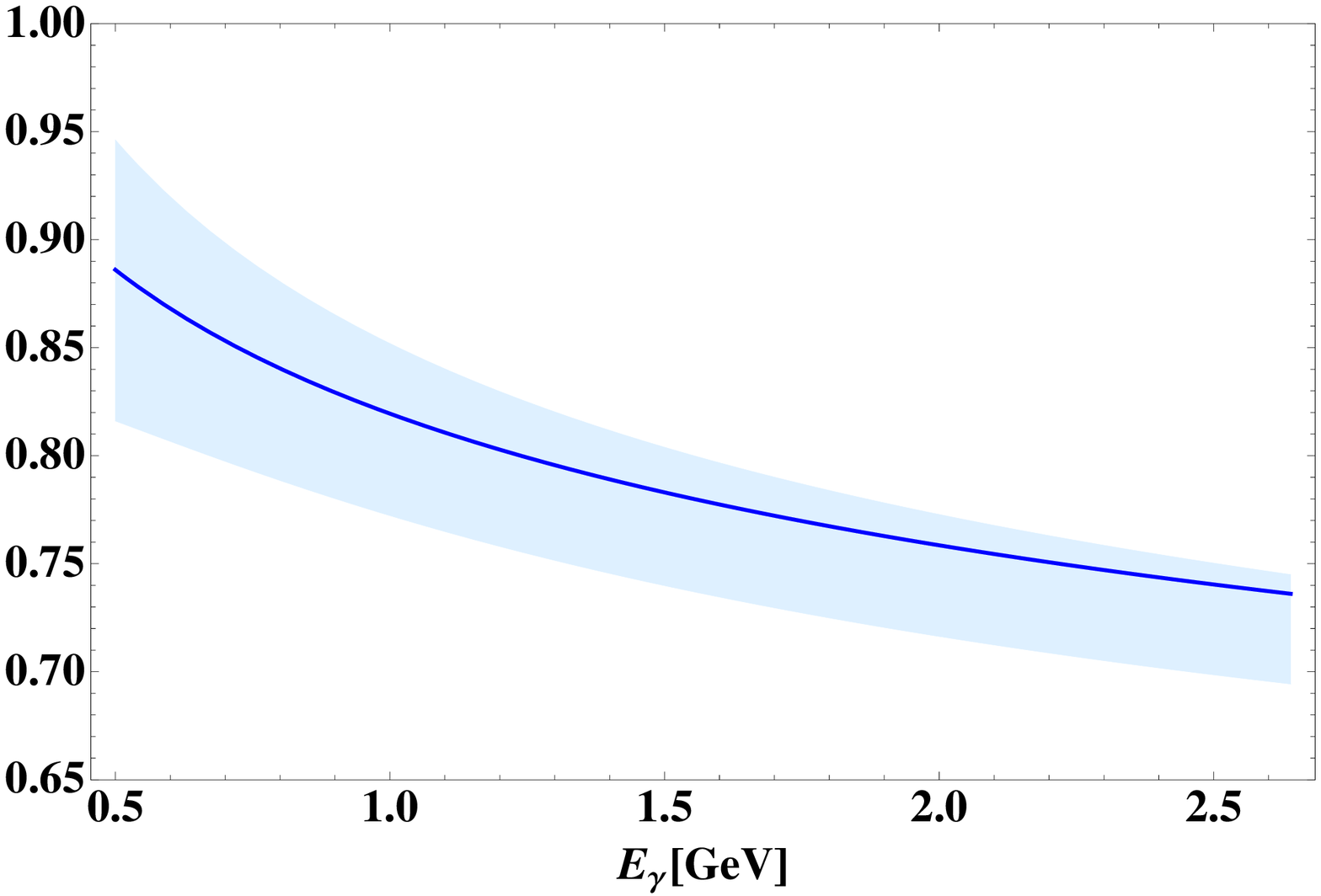}   
\end{center} 
\caption{Left: Hard-collinear scale dependence of the radiative  
correction factor $[\lambda_B(1\,\mbox{GeV})/\lambda_B(\mu)]\times  
R(E_\gamma,\mu)$ for $E_\gamma=2.0\,{\rm GeV}$. Right: Energy dependence 
with uncertainty band due to residual scale dependence.} 
\label{fig:scaledep} 
\end{figure}

The radiative corrections encoded in $R(E_\gamma,\mu)$ are important,  
reducing the leading-order amplitude by $20-25\%$. To judge the accuracy  
of the NLL computation, we show the residual dependence on the  
hard-collinear scale $\mu$ in Fig.~\ref{fig:scaledep}. We plot  
$[\lambda_B(1\,\mbox{GeV})/\lambda_B(\mu)]\times  
R(E_\gamma,\mu)$ for $E_\gamma=2.0\,{\rm GeV}$, which is the 
quantity that should be scale-independent,  if the radiative
corrections were known with infinite precision. The  
scale dependence of the prefactor follows from the evolution  
equation of the $B$-meson LCDA~\cite{Lange:2003ff} and is given  
by 
\begin{equation} 
\frac{\lambda_B(\mu_0)}{\lambda_B(\mu)} =  
1+\frac{\alpha_s(\mu_0) C_F}{4\pi}\ln\frac{\mu}{\mu_0} 
\left[2-2\ln\frac{\mu}{\mu_0} -4\sigma_1(\mu_0)\right]. 
\label{lambscaledep} 
\end{equation} 
Note that we do not sum logarithms of $\mu/\mu_0$,  
since $\mu_0=1\,\mbox{GeV}$, though formally a hadronic scale of few  
$\times \Lambda_{\rm QCD}$, is quite close to the hard-collinear  
scale $\mu \approx 1.5\,\mbox{GeV}$.  
In the numerical evaluation of $R(E_\gamma,\mu)$ we multiply out  
all $(1+\mbox{const.}\times\alpha_s)$ factors that originate from  
the NLO matching coefficients and evolution factors, but not the  
one from (\ref{lambscaledep}), and drop  
${\cal O}(\alpha_s^2)$ terms, which are beyond the NLL  
approximation. We also set the hard-matching scales to  
$\mu_{h1}=\mu_{h2}=m_b$.  
 
Fig.~\ref{fig:scaledep} (left panel)  
shows that the residual scale-dependence of  
the NLL approximation (solid line) is quite small. Recalling that  
$R(E_\gamma,\mu)$ equals 1 in the absence of any radiative correction,  
we see that the LL correction (dashed) is small; the main radiative  
effect arises from the NLO correction to the matching  
coefficients (\ref{hardPart}) and (\ref{hcPart}) rather than the  
summation of logarithms. However, comparing the NLL result  
to the unresummed NLO calculation (dotted, obtained from  
setting $\mu_{h1}=\mu_{h2}=\mu$), we note that renormalization  
group improvement stabilizes the scale-dependence at low $\mu$ and  
hence improves the accuracy of the result. An analysis of the  
residual dependence on the hard matching scales shows that it is  
of similar size as the hard-collinear scale dependence. 
  
The photon-energy dependence of the radiative correction factor  
is shown in the right panel of Fig.~\ref{fig:scaledep}.  
The shaded band represents the theoretical uncertainty estimated from  
varying the hard-collinear scale in the interval $[1,2]\,\mbox{GeV}$  
around the default value $\mu=1.5\,$GeV and the hard scales  
$\mu_{h1}=\mu_{h2}$ in $[m_b/2,2 m_b]$ around $m_b$. The uncertainties  
from each variation are added in quadrature. We conclude that radiative  
corrections reduce the $B\to\gamma\ell\nu$ amplitude over the entire energy  
range, and more significantly at high photon energies. 
 
\begin{figure}[t] 
\begin{center} 
\includegraphics[width=.66\textwidth]{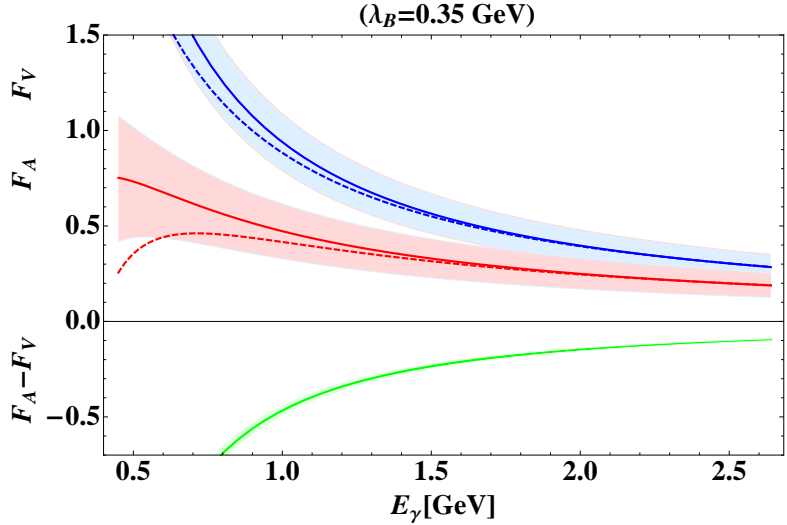}   
\end{center} 
\caption{Form factors $F_V(E_\gamma)$, $F_A(E_\gamma)$ and their  
difference. The bands show the total theoretical uncertainty, for  
fixed $\lambda_B(\mu_0)=0.35\,$GeV.} 
\label{fig:formfactors} 
\end{figure} 
 
The key quantities for the computation of differential decay  
distributions are the two form factors $F_V$, $F_A$ given  
in (\ref{ffs}). We display them in Fig.~\ref{fig:formfactors},  
which summarizes our main theoretical result. To obtain the form factors  
we need an ansatz for the size and energy dependence of the  
symmetry-conserving form factor $\xi(E_\gamma)$.  
The only information available is that it is a power correction  
of order $1/m_b$ to the leading term. We propose the form 
\begin{equation} 
   \xi(E_\gamma)= c \cdot \frac{f_B}{2 E_\gamma} \,, 
\end{equation} 
which features the same dependence on $E_\gamma$ as the leading  
term with $Q_u/\lambda_B$ replaced by $1/m_b$.  
The constant $c$ will be varied between $-1$ and $+1$.  
Fig.~\ref{fig:formfactors}  
displays $F_V$, $F_A$ and their difference including the theoretical  
uncertainty from adding in quadrature the scale uncertainty (as  
discussed above), the parameter $c$ and the input parameters from  
Tab.~\ref{InputData}, except $\lambda_B$. We do not include the  
$\lambda_B$ variation into the error here, since we intend to  
use $B\to\gamma\ell\nu$ to determine $\lambda_B$. How well this can  
be done depends on the theoretical uncertainty in $F_V$, $F_A$  
due to all other parameters. For comparison we also show in  
Fig.~\ref{fig:formfactors} the predicted form factors, when the  
hard scale $\mu_{h1}$ in the SCET matching coefficient is set to  
$2 E_\gamma$ (dashed lines). The difference to the standard  
choice $\mu_{h1}=m_b$ becomes significant only at very small  
photon energies. Since the factorization approach requires  
$2 E_\gamma\sim m_b$ the calculation of the form factors  
below photon energies of $1\,$GeV should certainly be considered  
unsafe. 
 
Recall that both form factors are exactly equal at leading order  
in the heavy-quark expansion. The difference between the two curves  
referring to $F_V$ and $F_A$ is therefore a direct measure of the  
magnitude of power corrections, which indeed rises at smaller  
photon energies, where the calculation breaks down. It is interesting  
to note that $F_V$ is predicted to rise faster than $F_A$ towards  
small energies, which is compatible with a pole dominance ansatz  
to model the low-energy regime~\cite{Becirevic:2009aq}. The  
uncertainties of $F_V$ and $F_A$ are highly correlated. This is seen  
explicitly when plotting the difference $F_V-F_A$ (lowest band in  
Fig.~\ref{fig:formfactors}), which has a very small uncertainty. In  
fact, from (\ref{ffs}) we obtain the definite prediction 
\begin{equation} 
F_A(E_\gamma) - F_V(E_\gamma) = \frac{f_B}{E_\gamma} 
\left[Q_\ell - \frac{Q_b m_B}{m_b} -  
\frac{Q_u m_B}{2 E_\gamma}  
\right]  
\label{ffdiff} 
\end{equation} 
up to corrections of order $\alpha_s$. Thus, the form-factor difference  
depends only on $f_B$. It would be very interesting to test this  
prediction of a power-suppressed effect experimentally. Looking at  
(\ref{doublediff}) we see that this can be done by selecting events  
with $x_\nu \approx 1$, i.e.~where the neutrino has nearly maximal  
energy and recoils against the lepton and the photon. Requiring a  
minimum separation angle between the lepton and the photon removes  
the non-radiative $B^-\to\ell \bar \nu$ contribution, which is, however,  
very small for the electron final state. Since $F_V-F_A$ is  
suppressed relative to $F_V+F_A$ the loss of statistics does not allow  
this test to be performed presently, but it should be within reach of  
the SuperB factories. 
 
As $F_{V/A} \propto 1/\lambda_B$ approximately, a measurement of the  
two form factors through $B\to\gamma\ell\nu$ can easily be turned into  
a determination of $\lambda_B$, within the uncertainties of the  
theoretical prediction shown in Fig.~\ref{fig:formfactors}. At present  
only upper limits exist on the branching fraction, resulting in  
lower bounds on $\lambda_B$ rather than a determination. We discuss  
the present limits and the dependence of partial branching  
fractions on $\lambda_B$ in the following.

\section{Bound on $\lambda_B$ from (partial)  
branching fractions}  
\label{Experiment:sec}  
 
\begin{figure}[t] 
\begin{center} 
\includegraphics[width=.48\textwidth]{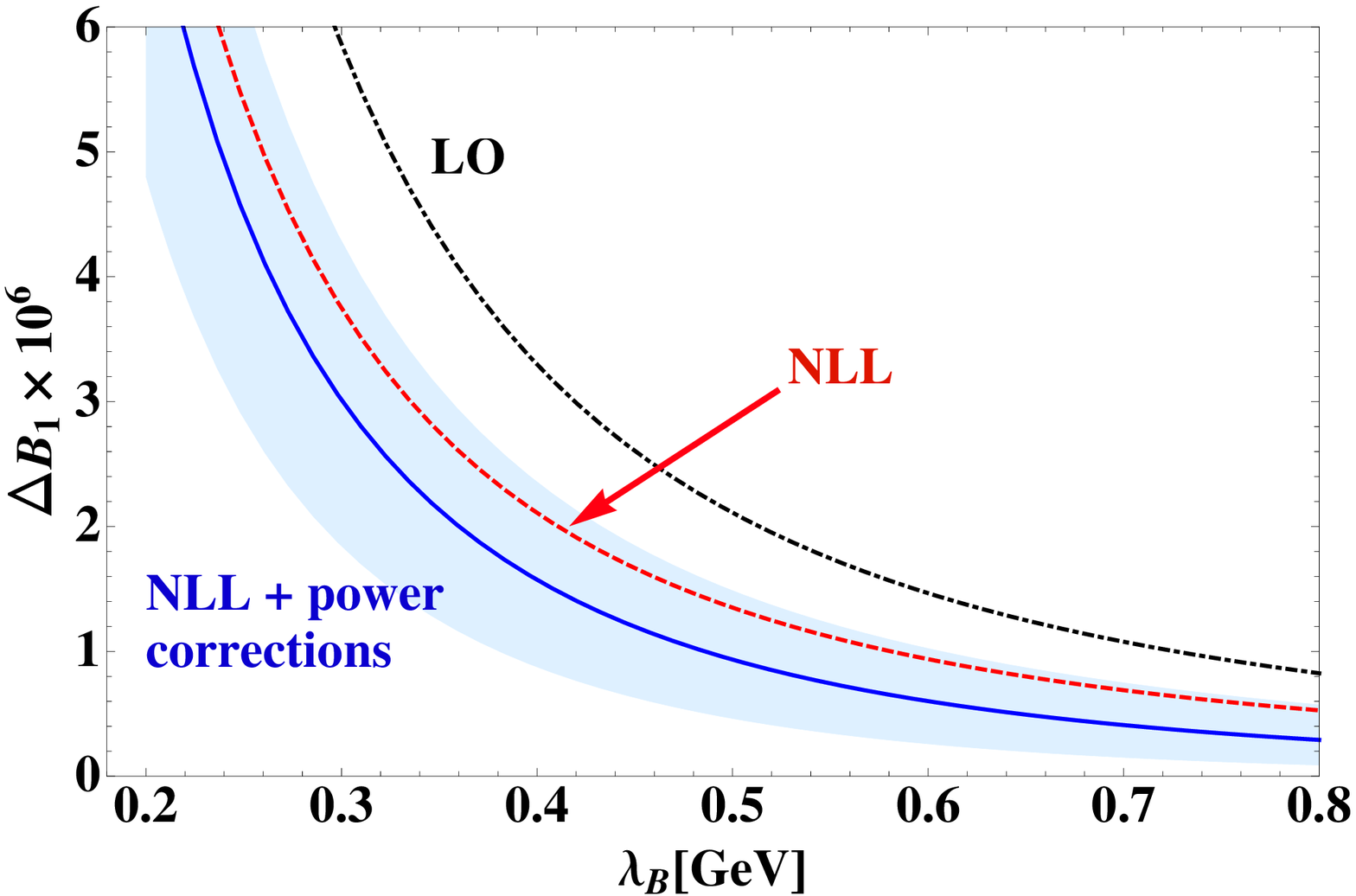}   
\includegraphics[width=.49\textwidth]{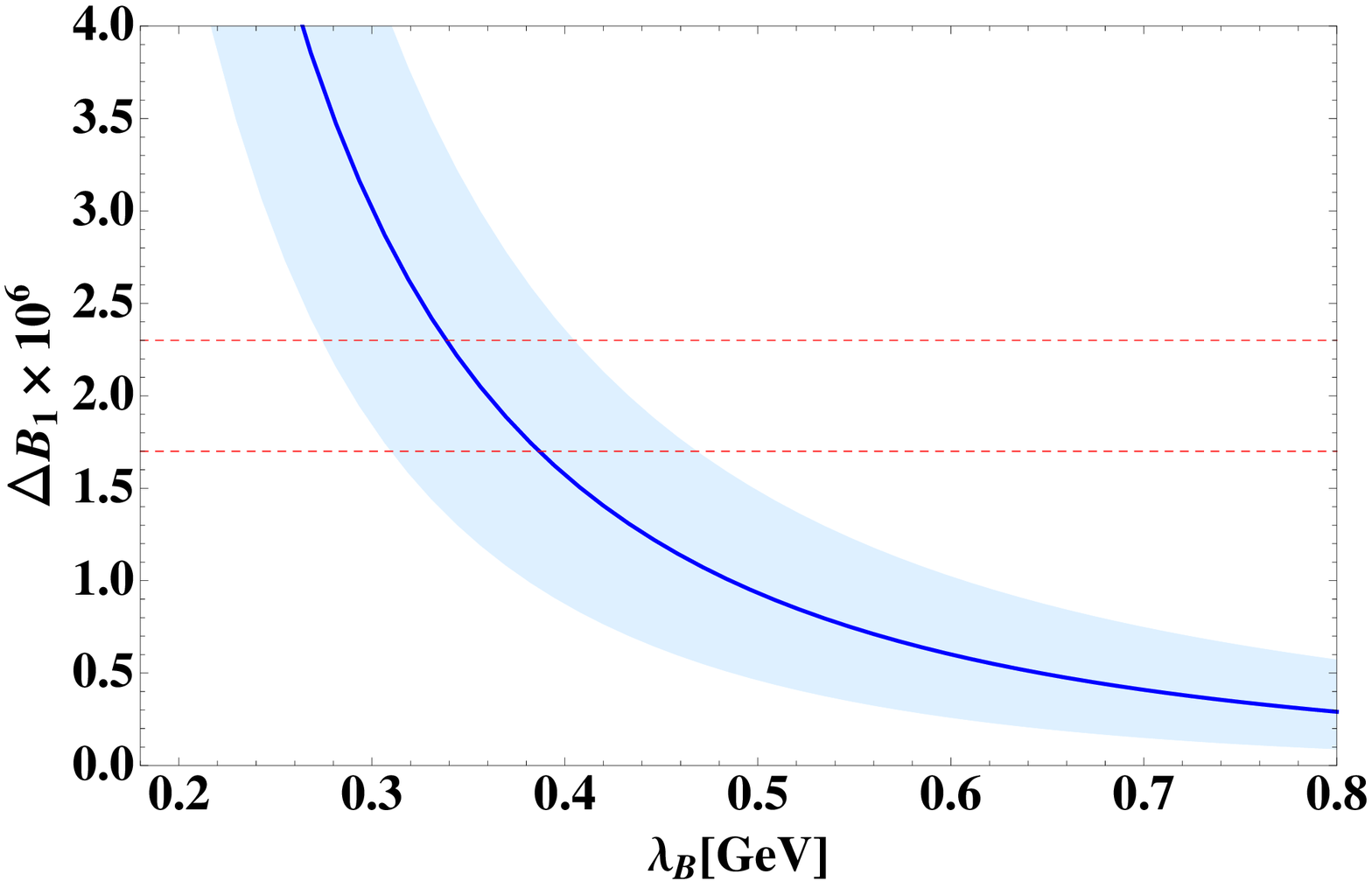}   
\end{center} 
\caption{Dependence of the  
partial branching fraction $\Delta \mathcal{B}_1$ on  
$\lambda_B$. Left: LO (dot-dashed), NLL without (with) power  
corrections (dotted, solid). Right: Theoretical prediction  
compared to BABAR limits~\cite{Aubert:2007yh}.}  
\label{fig:deltaB} 
\end{figure}

Experimental studies of the radiative leptonic decay have 
been performed by the CLEO Collaboration \cite{Browder:1996dt} 
and, more recently, by the BABAR Collaboration
\cite{Aubert:2007yh, Aubert:2009ya}.
The BABAR analyses make use of partial branching fractions  
\begin{equation} 
\Delta \mathcal{B}=\tau_{B_d} \int\limits_{\text{PS-Cuts}}  
\!\!\rd E_\gamma\;\rd E_\ell\;\; 
\frac{\rd ^2\Gamma}{\rd E_\gamma\;\rd E_\ell} \,. 
\end{equation} 
The first analysis \cite{Aubert:2007yh} employs the cuts 
$E_\ell \in (1.875 , 2.64)\,{\rm GeV}$,   
$E_\gamma \in ({0.45}, 2.35) \,{\rm GeV}$,  
$\cos\theta (\ell,\gamma)<-0.36$ in the cms frame of the $e^+ e^-$ collision,  
and quotes $\Delta \mathcal{B}_1 < 1.7 \,(2.3) \times 10^{-6}$ at  
90\% CL for flat priors on the amplitude (branching fraction). The  
second, published analysis~\cite{Aubert:2009ya} imposes only  
$E_\gamma<1\,\mbox{GeV}$ and yet finds the much weaker limit  
$\Delta \mathcal{B}_2 < 14\times 10^{-6}$. However, both analyses  
compare to a theoretical prediction that omits radiative corrections  
and contains an incorrect and numerically rather different  
expression for the $1/m_b$ power corrections. In Fig.~\ref{fig:deltaB},  
left panel, we show our prediction for $\Delta  \mathcal{B}_1$  
including uncertainties (solid, with band) for given $\lambda_B$,  
and compare it to the approximation without power corrections  
(NLL, dashed) and further omitting radiative corrections  
(LO, dot-dashed). Both effects together reduce $\Delta  \mathcal{B}_1$  
by more than a factor of two.\footnote{In our calculation we  
neglect the momentum $p_B\approx 0.3\,$GeV of the $B$ meson in the  
$e^+ e^-$ cms frame and apply the cuts directly in the $B$ rest frame. 
This approximation reproduces Eq.~(2) in \cite{Aubert:2007yh} to
excellent accuracy when adopting their theoretical input.}  
 
We first revisit the analysis of~\cite{Aubert:2007yh}. In this work  
$f_B=216\,$MeV and the rather large value $|V_{ub}|=0.00431$ are used,  
which further magnifies the theoretical prediction. In our analysis  
we adopt $|V_{ub}|_{\,\rm excl.}$ given in Tab.~\ref{InputData}, since  
all exclusive decays except $B\to\tau\nu$ tend to favor this smaller  
value. In Fig.~\ref{fig:deltaB} we show the two BABAR limits on  
$\Delta \mathcal{B}_1$ (straight lines). The lower limit on  
$\lambda_B\equiv \lambda_B(1\,\mbox{GeV})$  
follows from intersecting the lower end of the theoretical prediction  
with the straight lines. We then find that $\lambda_B > 310 \,(274)\, 
\mbox{MeV}$ at $90\%$ CL compared to $\lambda_B > 669 \,(591)\,\mbox{MeV}$ given  
in~\cite{Aubert:2007yh}. With the larger value of  
$|V_{ub}|_{\,\rm incl.}$ derived from inclusive semi-leptonic decays  
given in Tab.~\ref{InputData} we obtain  $\lambda_B > 372 \,(331)\, 
\mbox{MeV}$, which is still significantly smaller that the previous  
values, and illustrates the importance of radiative and power  
corrections. The definition of $\Delta \mathcal{B}_1$ is far from  
ideal from the theoretical point of view, since it includes photons  
with energies down to $0.45\,$GeV, where the theoretical prediction  
is not valid. The above limits should therefore be taken with a grain  
of salt. In this respect, $\Delta \mathcal{B}_2$ is somewhat  
better suited, but the weak experimental limit~\cite{Aubert:2009ya}  
results in the  
rather weak limit $\lambda_B > 115\,\mbox{MeV}$.  
 
\begin{figure}[t] 
\begin{center} 
\includegraphics[width=.66\textwidth]{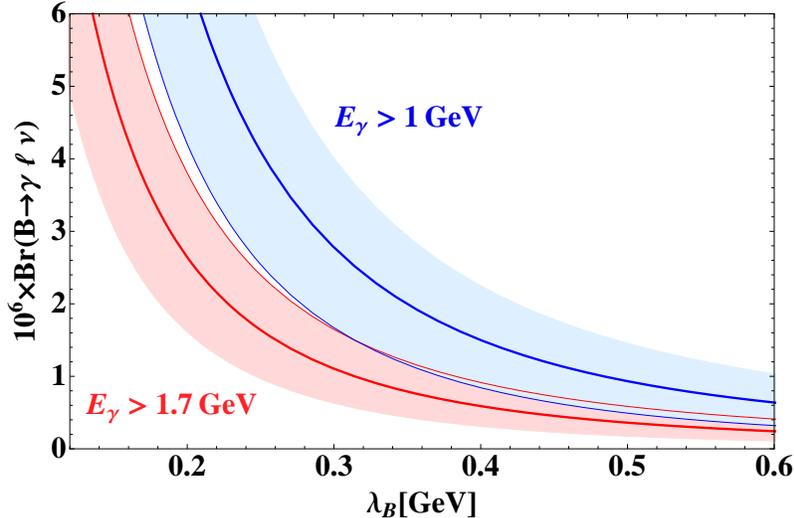}   
\end{center} 
\caption{The partial branching fractions   
$\mbox{Br}\,(B^-\!\to\gamma\ell\bar\nu, E_\gamma> E_{\rm cut})$  
for $E_{\rm cut}=1\,$GeV (upper band) and  
$1.7\,$GeV (lower band).} 
\label{fig:finalBR2} 
\end{figure}

We conclude that present data does not yet allow us to put significant  
constraints on $\lambda_B$. However, the theoretical prediction  
of the form factors is sufficiently accurate such that  
$B\to\gamma\ell\nu$ holds great promise for the future. In  
Fig.~\ref{fig:finalBR2} we show the inclusive branching fraction  
for a photon-energy cut $E_\gamma>1\,$GeV (upper band,  
equal to $\Delta \mathcal{B}_2$) and $E_\gamma>1.7\,$GeV, the  
latter being on more solid grounds theoretically. We see that  
a hypothetical measurement of $\mbox{Br}\,(B^-\to\gamma\ell\bar\nu,  
E_\gamma>1.7\,\mbox{GeV}) = 2.0 \times 10^{-6}$ with a 20\% error  
would constrain $\lambda_B$ to $[167,304]\,$MeV with a central  
value of $228\,$MeV.  
 
Can this be improved?  
The dominant theoretical errors arise from $\xi(E_\gamma)$, and the  
inverse-logarithmic moments $\sigma_1$, $\sigma_2$. It is hard to  
conceive of theoretical tools that would determine these quantities  
without providing $\lambda_B$ itself, rendering the present analysis  
superfluous. From (\ref{hcPart}) we see that $\sigma_1$ influences  
the shape of the normalized photon-energy spectrum. But this dependence  
is rather weak when the photon-energy cut is large enough to be solidly  
in the perturbative regime, making an extraction of $\sigma_1$  
difficult. We should mention though, that our error bands are based  
on rather conservative error ranges. For instance, we increased the  
error on $\sigma_1$ given in~\cite{Braun:2003wx} by a factor of 2.5,  
since this is the only attempt to estimate $\sigma_1$ up to now. 
  
\section{Conclusion}  
\label{conl:sec}  
 
We analyzed the radiative leptonic $B^-\to\gamma\ell\bar\nu$ decay  
with respect to its utility for determining the $B$ meson  
light-cone distribution amplitude, in particular its inverse  
moment, $\lambda_B$. We presented predictions for the form factors  
$F_V$, $F_A$ governing this decay, including  
for the first time radiative corrections and the leading-power  
corrections, and detailed uncertainty estimates. Corrections  
to the leading-order prediction reduce the branching fraction  
significantly. The BABAR upper limits on $B^-\to\gamma\ell\bar\nu$  
therefore presently do not allow to put stringent constraints on  
$\lambda_B$.  
We also showed that the power-suppressed difference of the two  
form factors can be predicted at leading order. 
The hundred-fold increase in statistics available to future  
$B$ factories therefore makes $B^-\to\gamma\ell\bar\nu$ an  
interesting process for determining $\lambda_B$ and testing  
the theory of power corrections in hard, exclusive $B$ decays. 
 
\subsubsection*{Acknowledgements}
We thank S.~J\"ager for helpful comments.
This work is supported in part by the DFG 
Sonder\-for\-schungs\-bereich/Trans\-regio~9 ``Computergest\"utzte 
Theoreti\-sche Teilchenphysik''.

\begin{appendix} 
\section{Renormalization group evolution factor} 
\label{appendix} 
 
The summation of formally large logarithms from the 
hard-to-hard-collinear scale ratio is accomplished by the renormalization 
group equation for the hard matching coefficients, or, equivalently  
the evolution factor $U(E_\gamma,\mu_{h1},\mu_{h2},\mu)=  
U_1(E_\gamma,\mu_{h1},\mu) U_2(\mu_{h2},\mu)^{-1}$. The first factor  
is associated with the running of the SCET current 
$\bar\xi W_c \gamma_{\mu_\perp} h_v$ and  
satisfies \cite{Bosch:2003fc} 
\begin{equation} 
\mu\frac{d}{d\mu} \,U_1(E_\gamma,\mu_h,\mu) =  
\left(\Gamma_{\rm cusp}(\alpha_s)\,\ln\frac{\mu}{2 E_\gamma} +  
\gamma(\alpha_s)\right) U_1(E_\gamma,\mu_h,\mu) 
\label{rge} 
\end{equation} 
with initial condition $U_1(\mu,\mu)=1$.  
We expand the anomalous dimension and the QCD beta-function  
according to  
\begin{equation} 
\gamma(\alpha_s) = \sum_{n=0} \gamma_n  
\left(\frac{\alpha_s}{4\pi}\right)^{n+1} 
\,\qquad 
\beta(\alpha_s) = \mu\frac{d\alpha_s}{d\mu} =  
-2\alpha_s  \sum_{n=0} \beta_n \left(\frac{\alpha_s}{4\pi}\right)^{n+1} 
\end{equation} 
(similarly for $\Gamma_{\rm cusp}$). The solution to (\ref{rge})  
is 
\begin{eqnarray} 
U_1(E_\gamma,\mu_h,\mu) &=&  
\exp\left(\,\int_{\alpha_s(\mu_h)}^{\alpha_s(\mu)} d\alpha_s\, 
\left[ 
\frac{\gamma(\alpha_s)}{\beta(\alpha_s)} +  
\frac{\Gamma_{\rm cusp}(\alpha_s)}{\beta(\alpha_s)}  
\left( 
\ln\frac{2 E_\gamma}{\mu_h} -  
\int_{\alpha_s(\mu_h)}^{\alpha_s}  
\frac{d\alpha_s^\prime}{\beta(\alpha_s^\prime)} 
\right) 
\right]\right) 
\nn \\[0.4cm] 
&& \hspace*{-2.2cm} 
=\,\exp\left(\, 
-\frac{\Gamma_0}{4\beta_0^2} \left( 
\frac{4\pi}{\alpha_s(\mu_h)}\left[\ln r-1+\frac{1}{r}\right]  
-\frac{\beta_1}{2 \beta_0} \,\ln^2 r 
+\left(\frac{\Gamma_1}{\Gamma_0}-\frac{\beta_1}{\beta_0}\right) 
\left[r-1-\ln r\right]\right) 
\right) 
\nn \\[0.2cm] 
&& \hspace*{-1.7cm} 
\times\,\left(\frac{2 E_\gamma}{\mu_h}\right)^ 
{-\frac{\Gamma_0}{2\beta_0} \ln r}  
 r^{-\frac{\gamma_0}{2\beta_0}}  
\times \Bigg[1 - \frac{\alpha_s(\mu_h)}{4\pi}\,\frac{\Gamma_0}{4\beta_0^2} 
\,\bigg(\frac{\Gamma_2}{2\Gamma_0} \left[1-r\right]^2  
+\frac{\beta_2}{2\beta_0} \left[1-r^2+2 \ln r\right]  
\nn \\[0.2cm] 
&& \hspace*{-0.7cm} 
-\,\frac{\Gamma_1\beta_1}{2\Gamma_0\beta_0}  
\left[3-4 r+r^2+2 r \ln r\right]  
+\frac{\beta_1^2}{2\beta_0^2} \left[1-r\right]\left[1-r-2\ln r\right] 
\bigg) 
\nn \\[0.2cm] 
&& \hspace*{-0.7cm} 
+\,\frac{\alpha_s(\mu_h)}{4\pi}\left( 
\ln\frac{2E_\gamma}{\mu_h}  
\left(\frac{\Gamma_1}{2\beta_0}-\frac{\Gamma_0\beta_1}{2\beta_0^2}\right) 
+\frac{\gamma_1}{2\beta_0}-\frac{\gamma_0\beta_1}{2\beta_0^2}\right) 
\left[1-r\right] + {\cal O}(\alpha_s^2)\Bigg]  
\label{U1} 
\end{eqnarray} 
with $r = \alpha_s(\mu)/\alpha_s(\mu_h)$. After the second equality  
the exact solution has been expanded to NLL. At this order the cusp  
anomalous dimension enters at the three-loop order 
\cite{Moch:2004pa}. Its series coefficients are  
\begin{eqnarray} 
&& \Gamma_0=4 C_F,  
\qquad 
\Gamma_1=C_F\left[\frac{268}{3}-4\pi^2-\frac{40}{9} n_l\right], 
\\ 
&& 
\Gamma_2 = C_F\left[1470-\frac{536\pi^2}{3}+\frac{44\pi^4}{5}+264\zeta(3) 
+n_l\left(-\frac{1276}{9}+\frac{80\pi^2}{9}-\frac{208}{3}\zeta(3)\right) 
-\frac{16}{27}n_l^2\right], 
\nn 
\end{eqnarray} 
where $n_l=4$ is the number of light fermion flavours (the charm quark 
is treated as massless), and $C_F=4/3$ 
the quadratic Casimir of the fundamental SU(3) representation. The remaining  
anomalous dimension of the SCET heavy-light current is needed  
at two loops, and given by 
\begin{equation} 
\gamma_0=-5 C_F,\qquad  
\gamma_1 = C_F\left[-\frac{1585}{18}-\frac{5\pi^2}{6}+34\zeta(3)+ 
n_l\left(\frac{125}{27}+\frac{\pi^2}{3}\right)\right]\,. 
\end{equation} 
The two-loop expression is given explicitly  
in \cite{Asatrian:2008uk,Bell:2008ws} confirming an earlier conjecture 
\cite{Neubert:2004dd}. 
 
The second evolution factor $U_2(\mu_{h2},\mu)$ arises from the matching  
of the $B$ meson decay constants in QCD and heavy-quark effective  
theory (HQET). Its expression follows from the ones given above by  
setting the cusp anomalous dimension to zero, and by replacing  
$\gamma_i$ by the anomalous dimension of the heavy-light current  
in HQET, $\gamma_{i,\rm hl}$,  
given to two loops by \cite{Ji:1991pr,Broadhurst:1991fz} 
\begin{equation} 
\gamma_{0,\rm hl}=-3 C_F,\qquad  
\gamma_{1,\rm hl} = C_F\left[-\frac{127}{6}-\frac{14\pi^2}{9} 
+\frac{5}{3} n_l\right]\,. 
\end{equation} 
 
The three-loop evolution of the strong coupling in the  
$\overline{\rm MS}$ scheme is computed from  
\begin{eqnarray} 
\alpha_s(\mu) &=& \frac{4\pi}{\beta_0\ln(\mu^2/\Lambda^2)} 
\Bigg[1-\frac{\beta_1}{\beta_0^2}  
\frac{\ln\ln(\mu^2/\Lambda^2)}{\ln(\mu^2/\Lambda^2)}  
+ \frac{\beta_1^2}{\beta_0^4 \ln^2(\mu^2/\Lambda^2)} 
\nn \\ 
&&\times\, 
\left(\left(\ln\ln(\mu^2/\Lambda^2)-\frac{1}{2}\right)^2+ 
\frac{\beta_2\beta_0}{\beta_1^2}-\frac{5}{4}\right) 
\Bigg] 
\label{asrun} 
\end{eqnarray} 
with 
\begin{equation} 
\beta_0= 11-\frac{2 n_l}{3},\qquad  
\beta_1 = 102 -\frac{38 n_l}{3} \, \qquad 
\beta_2 = \frac{2857}{2} - \frac{5033}{18} n_l + \frac{325}{54} n_l^2. 
\end{equation}

\end{appendix}

\end{document}